\documentclass[twocolumn,showpacs,showkeys,pre,floatfix]{revtex4}

\usepackage{bm}
\usepackage{graphicx}
\usepackage{color}
\usepackage[usenames,dvipsnames]{xcolor}
\usepackage{amsmath, amsthm, amssymb}
\usepackage{theorem}
\usepackage{ascmac}

\begin{document}

\title{Thermodynamic Characterization of Synchronization-Optimized Oscillator-Networks}

\author{Tatsuo Yanagita}
\affiliation{Department of Engineering Science, Osaka Electro-Communication University, Neyagawa 572-8530, Japan}
\email{yanagita@isc.osakac.ac.jp}
\homepage{http://phenomath.osakac.ac.jp}
\author{Takashi Ichinomiya}
\affiliation{Department of Biomedical Informatics, Gifu University School of Medicine,
1-1 Yanagido, Gifu, Gifu 501-1194, Japan}

\begin{abstract}
We consider a canonical ensemble of synchronization-optimized networks of identical oscillators under external noise.
By performing a Markov chain Monte Carlo (MCMC) simulation using the Kirchhoff index, i.e., the sum of the inverse eigenvalues of the Laplacian matrix (as a graph Hamiltonian of the network), we construct more than 1,000 different synchronization-optimized networks.
We then show that the transition from star to core-periphery structure depends on the connectivity of the network, and is characterized by the node degree variance of the synchronization-optimized ensemble.
We find that thermodynamic properties such as heat capacity show anomalies for sparse networks.
\end{abstract}
\date{\today }

\maketitle

\newcommand{\singlefiguresize}{0.85\columnwidth}
\newcommand{\doublefiguresize}{1.85\columnwidth}


\section{Introduction}

In the past decade, studies of complex
networks consisting of dynamical elements involved in a
set of interactions \cite{RevModPhys.74.47,Boccaletti2006175} have attracted considerable interest. 
Particular attention has been paid to problems of synchronization in
network-organized oscillator systems \cite{Manrubia04,Arenas08}. 
Synchronization phenomena are ubiquitous in various fields of science
and play an especially important
role in the functioning of living systems \cite{Kurths01}.
In this research, emphasis has been placed on understanding
the relationship between the topological structure of a network and its
 collective synchronous behavior \cite{Boccaletti2006175}, and these studies
 have revealed the synchronization properties of systems
formed by phase oscillators on static complex networks, such as small-world
\cite{Hong02} and scale-free networks
\cite{Ichinomiya04,Lee05}.
In addition, this research has also shown that the ability of a network
to give rise to synchronous behavior can be greatly enhanced by exploiting
the topological structure emerging from the growth processes
\cite{PhysRevE.71.016116,PhysRevLett.94.138701}.
However, a full understanding of how network topology affects
 the synchronization of specific dynamical units is still lacking.

Application of evolutionary learning is one possible approach to
constructing networks with prescribed dynamical properties. On this issue, 
several methods have been explored, where network structure has been
modified in response to selection pressure via learning algorithms
in such a way that the system
evolved towards a specified goal
\cite{Mikhailov02,Moyano01,yanagita10,yanagita12}. 
In previous studies, we have applied the Markov chain Monte Carlo (MCMC)
method with replica exchange to design synchronization-optimized
networks \cite{yanagita10,yanagita12}. In these studies, we constructed
large ensembles of optimal networks for synchronization, which was
evaluated by the Kuramoto order parameter, and analyzed their common
statistical properties. For a network of heterogeneous phase
oscillators, we found the transition from the linear to
bipartite-like networks by increasing the number of
links \cite{yanagita10}.
For a network of identical phase oscillators under uncommon
noise, the most optimal network is a star-like structure when the number
of links is small, while an interlaced structure is preferable when the number of links is  large.
Unfortunately, in these studies we succeeded in achieving an optimized network only
 when the network size was small, and we also
failed to analyze the thermodynamic properties of the network ensemble in detail, due to the computational cost of calculating the Kuramoto
 order parameter.

In this paper, we consider a large network consisting of identical oscillators with uncommon noise, and present the structure  and  statistical properties of the optimized network.
To reduce the computational cost of evaluating synchronization performance,
we employ the sum of the inverse eigenvalues of the Laplacian matrix as
 the synchronization performance index. It is easily
shown that this value indicates the synchronization strength, under
the assumption that the connection is non-directional and that the noise
acting on each oscillator is sufficiently small.
The computational cost of obtaining the eigenvalues of the Laplacian is much smaller
than that of calculating the Kuramoto order parameter, and  we can perform MCMC to
 optimize the large oscillator network.

To be precise, the MCMC method used here is not an optimization, but rather samples networks from a given probability distribution. In this approach, we create an ensemble of networks that obey the
canonical distribution, in which the ``energy'' is defined by the
synchronization performance. Therefore MCMC enables us to study the
thermodynamic properties of this ensemble of optimized networks. The investigation of the thermodynamic properties will  give us information
  not only  on the most optimized network, but also on the second, third, or subsequent ``most 
  optimized" configurations. 
 In addition, thermal properties are used to characterize constraint-satisfaction problems such as $N$-queens and Latin problems
\cite{Hukushima200277,PhysRevE.79.016703}. Therefore, we investigate
the thermodynamic properties of these ensembles in this paper.

The paper is organized as follows. In Sec.~\ref{sec:model}, we introduce
 a model of identical phase oscillators occupying the nodes of a
 symmetrically coupled network, and define the synchronization
 measure as the sum of the inverse of the eigenvalues of the Laplacian matrix. 
The sampling method is also introduced in this section. 
Then, construction of the optimized networks and their thermodynamic
 characterization are performed in Sec.~\ref{sec:numerical} while, finally, the results are discussed in Sec.~\ref{sec:summary}.


\section{The Model and the Method}
\label{sec:model}

In this paper, we consider the effect of uncommon noise on the
synchronization of identical oscillators. The model equations are
\begin{equation}
\frac{d \theta_i}{dt}=\omega_0+\frac{\epsilon}{N} \sum_{j=1}^{N}w_{i,j}\sin(\theta_j-\theta_i)+\xi_i(t),
\label{eq:model}
\end{equation}
where $\epsilon$ is coupling strength and $\xi_i(t)$ represents independent
white noise, such that $\langle \xi_i(t) \rangle=0$ and $\langle
\xi_i(t) \xi_j(t')\rangle=D \delta_{i,j} \delta(t-t')$. $N$ and $\omega_0$ 
 are the number of oscillators and the natural frequency, respectively. $\theta_i$, the phase of the oscillator at node $i$, is between 0 and $2\pi$.
Interactions between the oscillators are specified by the adjacent
matrix $\mathbf{w}$, defined as $w_{i,j} = 1$ if there is a
connection between node $i$ and $j$, and $w_{i,j} = 0$ otherwise. In the
following, we only investigate non-directional networks, for which
$w_{i,j}=w_{j,i}$.

Because the rotation frequencies of all the oscillators are the same, we
can always go into the rotational frame $\theta_i \mapsto
\theta_i-\omega_0 t$ and, thus, eliminate the
term with $\omega_0$.
Hence, without any loss of generality, we can set $\omega_0=0$ in Eq.~(\ref{eq:model}).
It is known that the model shows the transition between the synchronized and
desynchronized state by changing the ratio of the coupling strength to the noise intensity \cite{Mikhailov06}.


In a previous study \cite{yanagita10,yanagita12}, in order to measure the
  degree of synchronization, we numerically integrated differential
  equations and calculated the
  long time average of the Kuramoto order parameter,
\begin{equation}
R=\lim_{T \rightarrow \infty}\int_0^{T}\frac{1}{N} \left| \sum_{i=1}^{N}\exp (\mathbf{i}\theta _{i}) \right| dt.
\end{equation}%
However, this numerical integration has a large computational cost,
which makes  estimation of the synchronization performance of a large
network difficult.
In order to overcome this difficulty, we estimate the synchronization
performance theoretically by assuming that the connection is symmetric and
the noise intensity is sufficiently small \cite{Korniss07}.
If there is no noise, all oscillators are completely synchronized, i.e.,
$\theta_i = \Theta$ for all $i$. If the noise is sufficiently small however, we can
linearize the interaction term, $\sin(\theta_j-\theta_i)$, as
$\theta_j-\theta_i$, and Eq.(\ref{eq:model}) is approximated by
\begin{equation}
 \frac{d \theta_i}{dt}=\frac{\epsilon}{N} \sum_{j=1}^{N}L_{i,j}\theta_j+\xi_i(t).\label{131240_30Jul14}
\end{equation} 
\begin{equation}
L_{i,j}=
  \begin{cases}
w_{i,j}\ , &(i\neq j) \\
\displaystyle {-\sum^{N}_{j=1}w_{i,j}}\ . & (i=j)
  \end{cases}
\end{equation}%
The static distribution of Eq. (\ref{131240_30Jul14}) is given by the multivariate Gaussian distribution.
Using the eigenvalues of the Laplacian matrix,
\[
0=\lambda_0 \geq \lambda_1 \geq \lambda_2 \geq \cdots \geq \lambda_{N-1},
\]
the Kuramoto order parameter of the system is given by
\begin{equation}
R =1+\frac{\epsilon}{4N} \sum_{i=1}^{N-1} \frac{1}{\lambda_i}.
\end{equation}%
Therefore, the sum of the inverse of the eigenvalues, $\Lambda$, where 
\begin{equation}
\Lambda=-\sum_{i=1}^{N-1}\frac{1}{\lambda_i}\ ,
\label{eq:sum}
\end{equation}
determines the synchronization performance. Hence, a network having a smaller
$\Lambda$ value shows better synchronization performance. 
We note that $\Lambda$ is related to the Kirchhoff index, which is a
 structure-descriptor of a molecular graph, and is also related to the
 effective electrical resistance in a resistive electrical network
 \cite{Babic02,Zhou09}. We employ $\Lambda$ as the synchronization performance indicator in this paper.

Next, we describe the MCMC method applied in the numerical simulations.
In this method, we randomly generate a connection network and estimate
the synchronization performance as indicated by $\Lambda$ (defined by Eq.~(\ref{eq:sum})).
It should be noted that estimation of the synchronization
performance based on $\Lambda$ is only valid for connected networks.
 In a case in which the network is disconnected, $\lambda_1=0$ and $\Lambda$
  diverges. However, allowing disconnected networks as stepping stones will 
  enhance the performance of the MC sampling. In order to include disconnected states in the form of stepping stones, we define
the energy, $E(\mathbf{w})$, as
\begin{equation}
E(\mathbf{w})=
\begin{cases}
	\Lambda\ , & (\lambda_1 \neq 0)\\
	\Lambda_{\max}+\sum_{i=1}^{N} \delta_{0,\lambda_i}\ , & (\lambda_1=0)
\end{cases}
\label{eq:hamiltonian}
\end{equation}
where $\delta_{i,j}$ is the Kronecker delta.
When $\lambda_1 \neq 0$, the network is connected, and we use $\Lambda$ itself as a synchronization indicator, in which smaller $E(\mathbf{w})$ 
shows better synchronization performance. 
If a network is disconnected, we introduce synchronization performance with a ``penalty'', as shown in the second line of Eq.~(\ref{eq:hamiltonian}).
The first term, $\Lambda_{\max}$, is the maximum of $\Lambda$ among the connected networks, and is given by $\Lambda$ when the network is a chain, i.e., $\Lambda_{\max}=\Lambda_{\mathrm{chain}}=(N^2-1)/6$, where $N$ is the number of nodes.

Through the function $E(\mathbf{w})$, we sample networks from the canonical ensemble, i.e.,
\begin{equation}
p(\mathbf{w})=\frac{\exp[-\beta E(\mathbf{w})]}{\sum_{\mathbf{w}} \exp[-\beta E(\mathbf{w})]},
\label{eq:canonical}
\end{equation}
borrowed from statistical mechanics. This sampling is carried out by the MCMC
 method, which has previously been applied to several dynamical systems \cite{Cho94,Bolhuis98,Vlugt00,Kawasaki05,Sasa06,Giardin06,Tailleur07,yanagita09,yanagita10,yanagita12}. 
The application of the canonical ensemble to a network is also considered in \cite{PhysRevE.70.066117,sym3010001}, in which $E$ is called the graph Hamiltonian.
In this work, we use the Replica Exchange Monte Carlo (REMC) algorithm \cite{Hukushima96,Iba01,Janke08} to sample synchronization-optimized networks.
It should be noted that disconnected networks are sampled using the graph Hamiltonian defined above, although we discard these networks for the calculation of the statistical properties discussed below.
Such a sampling enhances the mixing in the REMC through bridges (or paths) between networks with higher synchronization performance, for which disconnected networks are used as stepping stones \cite{Kikuchi99}.
Since the synchronization performance can be easily improved by
increasing the number of links, we consider the synchronization
performance for given networks with a fixed number of links.

In this work, we also investigate the density of states and other
thermodynamic properties.
Here, we briefly describe the method used to obtain these properties.
The density of states, $g(E)$, and the thermodynamic quantities can be calculated by WHAM.
First, $g(E)$ satisfies the equation
\begin{equation}
\exp(-\beta F_{\beta})=\sum_{E} g(E) \exp(-\beta E),
 \end{equation}
where $F_{\beta}$ is the free energy, defined by
\begin{equation}
\exp(-\beta F_{\beta})=Z(\beta)=\sum_{\mathbf{w}}  \exp(-\beta E(\mathbf{w})).
\end{equation}
The density of states can be estimated as
\begin{equation}
g(E)=\frac{H_{\beta}(E)}{\exp[F_\beta-\beta E]}\ ,
\end{equation}
by using the histogram of $E$, i.e., $H_{\beta}(E)$.
In REMC, since we have many histograms, $H_{\beta_m} \; (m=0, \cdots, M)$, obtained from many replicas with different inverse temperatures, $\beta_m$, a precise density estimate can be made by as follows, where
\begin{eqnarray}
g(E)&=&\frac{\sum_{m=1}^{M} H_{\beta_m}(E)}{\sum_{m=1}^{M} \exp[F_{\beta_m}-\beta_m E]}\ ,\\
\exp(-F_{\beta_m}) &=& \sum_{E}  g(E) \exp(-\beta_m E).
\end{eqnarray}
Here, $g(E)$ and $F_{\beta_m}$ can be obtained by satisfying the above equations as self-consistent solutions. These solutions can be easily obtained by iteration.

Using the density of states, we can calculate the thermodynamic average of any function of the graph, $X(\mathbf{w})$, by
\begin{equation}
\langle X \rangle_{\beta}=\frac{1}{L}\sum_{i=1}^{L} X(\mathbf{w}_i) g[E(\mathbf{w}_i)] \exp[-\beta E(\mathbf{w}_i)],
\label{eq:thermodynamic_average}
\end{equation}
where the sum is taken over $L$ sampled, connected networks.
Analogous to statistical physics, the ``heat capacity'', $c(\beta)$, can be expressed as fluctuation of $E$ with finite temperature, i.e.,
\begin{equation}
c(\beta) = \beta^2\left( \langle E^2 \rangle_\beta - \langle  E \rangle _\beta^2 \right),
\label{eq:heat_capacity}
\end{equation}
where
\begin{equation}
\langle E \rangle_\beta  = \sum_{E}  g(E) E \exp(-\beta E)\ . 
\end{equation}
is calculated through $g(E)$.

\section{Numerical analysis}
\label{sec:numerical}

To determine the synchronization degree of a given network at each iteration
step of the MCMC procedure, we calculated the eigenvalues of the Laplacian matrix using the linear algebra package, LAPACK \footnote{http://www.netlib.org/lapack/}.
Oscillator ensembles of sizes $N=10,15, 20, 25, 30,$ and $100$ were
considered.
Using the sum of the inverse eigenvalues, $\Lambda$, graphs were sampled using the REMC method. 
In parallel, evolution of $M+1$ replicas with inverse temperatures, $\beta _{m}=\kappa^m-1 \;\;
(m=0,1,\dots ,M)$, was performed mainly with $M=23$ and $\kappa =1.2$. 
After each 10 Monte Carlo steps (mcs), the performances of a randomly
chosen pair of replicas were compared and exchanged \cite{yanagita10,
yanagita12}. 
For display and statistical analysis, sampling at every 10,000 mcs, after a transient of $5,000$ mcs, was undertaken.

For convenience in the later discussion, we introduce the
 connectivity of a network with $N$ nodes and $K$
 links as $p=\frac{2(K-N+1)}{(N-1)(N-2)}$.
When the number of links is the minimum for a connected network, i.e.,
$K=N-1$, then $p=0$ and, when the network is fully connected, $p=1$.

\subsection{Architectures of synchronization-optimized networks}

\begin{figure}[tbp]
\begin{center}
 \resizebox{\singlefiguresize}{!}{\includegraphics{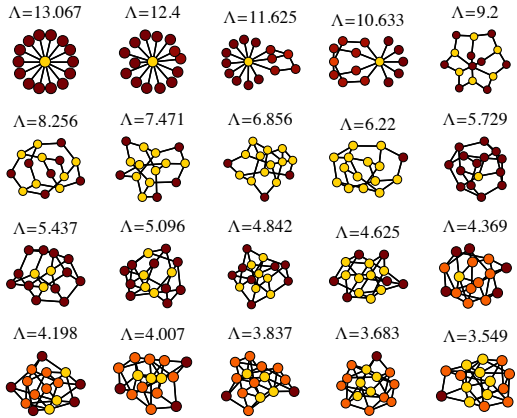}}
\end{center}
\vspace{-5mm}
\caption{(Color online). Most synchronization-optimized networks in order of increasing number of links.
The number of links of the network increases from $K=N-1,N,N+1,\cdots,N+18$, from the top left to the bottom right.
The nodes are colored according to their degree, and the lighter nodes indicate oscillators with a relatively higher degree in the network.
The parameters are $N=15, M=24$, and $\kappa=1.15$.}
\label{fig:graph_min}
\end{figure}

When $p$ equals zero, the most synchronization-optimized network
has a star structure (the star exhibits
 the best synchronization performance for $K=N-1$), as shown by the top-left
 graph in Fig.~\ref{fig:graph_min}. 
In this figure, the most synchronization-optimized networks are listed
 in order of increasing link number (connectivity).
The nodes are colored according to their degree, and the brighter nodes have
relatively higher degrees.
The value of $\Lambda$ is shown at the top of the each graph, and it is clear that the synchronization performance increases (i.e., $\Lambda$ decreases) with an increase in the number of links.
The star, which shows the best synchronization performance in networks with $N-1$ links, is inhomogeneous in the sense that the degree of the center node is $N-1$ and the degree of the other nodes is one.
The most optimized network transforms to a homogeneous network with an
 increase in the number of links.
For example, all of the synchronization-optimized networks in the second
row only have nodes with degrees between 2 and 4. 
In particular, the graph in the second row (the second from the right
in Fig.~\ref{fig:graph_min}), is the network in which all but one node have degree three.
When the connectivity is further increased, the core oscillators emerge and the network develops a more complex organization, forming a shallow tree (see the bottom-right graph in Fig.~\ref{fig:graph_min}).

\begin{figure}[tbp]
\begin{center}
\resizebox{\singlefiguresize}{!}{\includegraphics{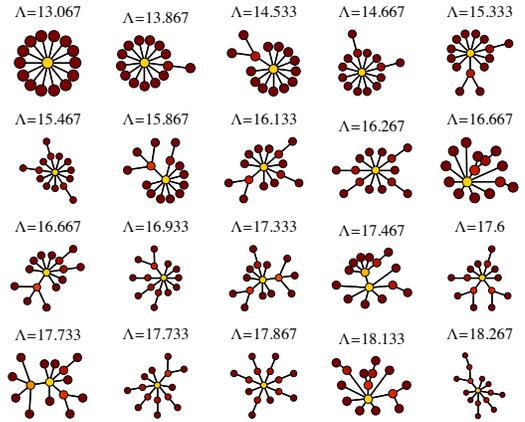}}
\end{center}
\vspace{-5mm}
\caption{(Color online). Series of graphs in order of increasing $\Lambda$ with $N=15$, $K=14$, and $p=0.0$.
The other parameters are the same as in Fig.~\ref{fig:graph_min}}.
\label{fig:typical_graph1}
\end{figure}

\begin{figure}[tbp]
\begin{center}
\resizebox{\singlefiguresize}{!}{\includegraphics{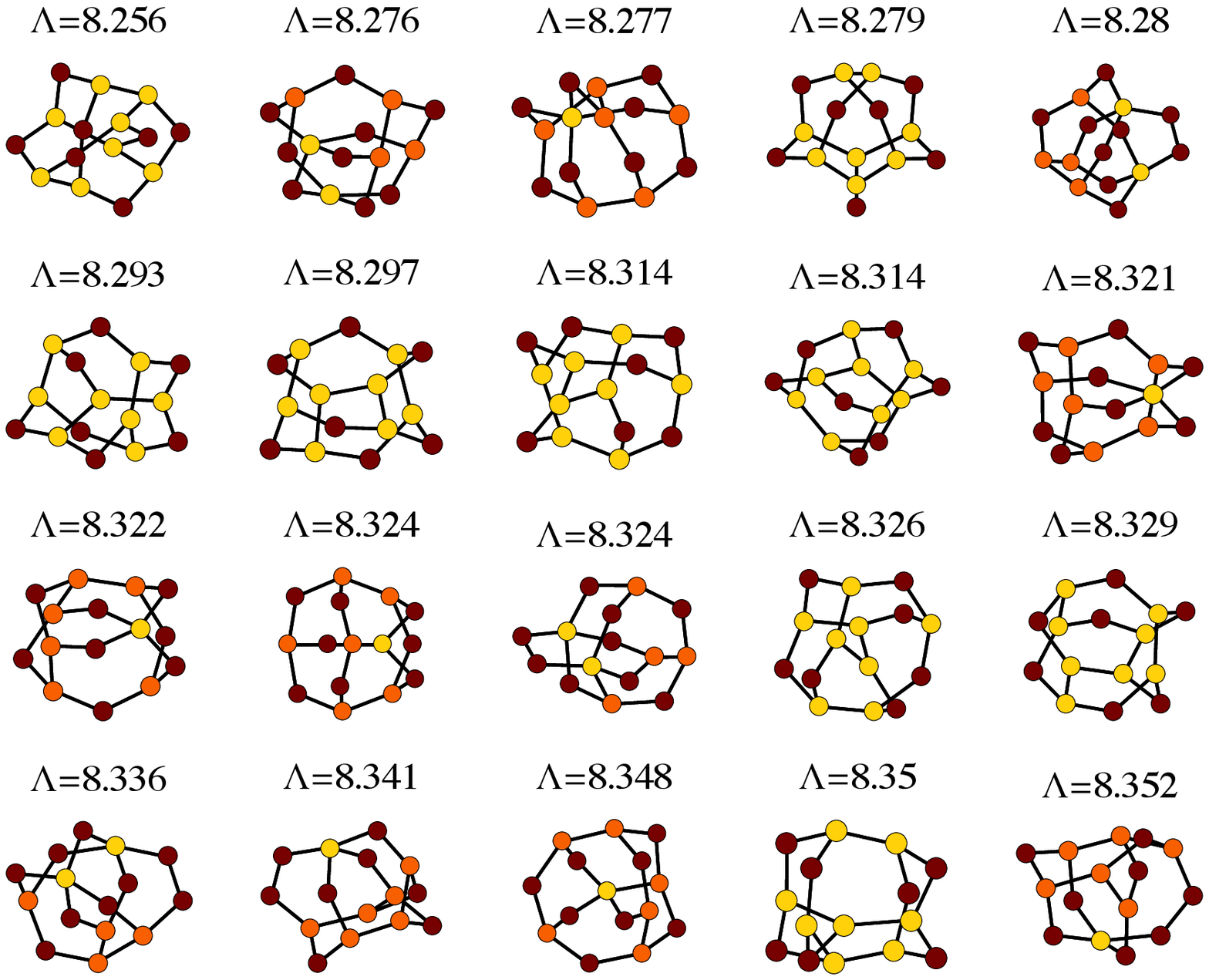}}
\end{center}
\vspace{-5mm}
\caption{(Color online). Series of graphs in order of increasing $\Lambda$ with $N=15$, $K=19$, and $p=0.0549451$.
The other parameters are the same as in Fig.~\ref{fig:graph_min}}.
\label{fig:typical_graph2}
\end{figure}

\begin{figure}[tbp]
\begin{center}
\resizebox{\singlefiguresize}{!}{\includegraphics{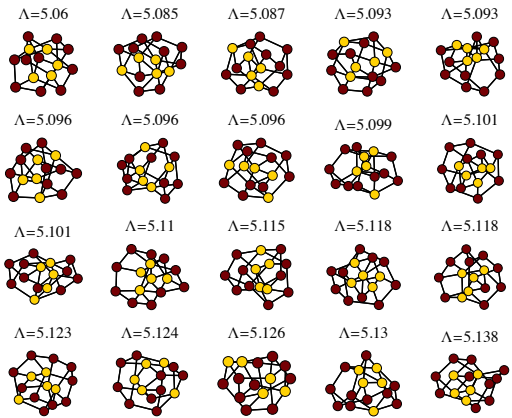}}
\end{center}
\vspace{-5mm}
\caption{(Color online). Series of graphs in order of increasing $\Lambda$ with  $N=15$ and $K=25$, and $p=0.120879$.
The other parameters are the same as in Fig.~\ref{fig:graph_min}}.
\label{fig:typical_graph3}
\end{figure}

To investigate the structure of a highly optimized network, we
investigate the structure of a network with small $\Lambda$.
In Fig.~\ref{fig:typical_graph1}, Fig.~\ref{fig:typical_graph2}, and
Fig.~\ref{fig:typical_graph3}, the series of graphs are listed in order of
increasing $\Lambda$ for $K=14,17,$ and $25$, respectively, with $N=15$.
Since the graph only develops a tree structure for $p=0$, as shown in
Fig.~\ref{fig:typical_graph1}, the most synchronization-optimized
network, i.e,  the star (top-left graph), becomes a complicated tree
as $\Lambda$ increases, and finally becomes a chain with the worst
synchronization performance.
When the number of links is slightly larger than $N-1$, the most optimized
network has a mesh structure, as shown in the top-left of Fig.~\ref{fig:typical_graph2}. 
In this case, the degree is almost homogeneous.
As the synchronization performance decreases, longer loops are more evident in the network, while the degree distribution does not show a significant difference to that of the most optimized network.
For higher connectivity, the synchronization performance depends only slightly
on the structure of the network (see the value of $\Lambda$ in
Fig.~\ref{fig:typical_graph3}), and significant differences
in the network structures are not apparent.

\begin{figure*}[tbp]
\begin{center}
\resizebox{\doublefiguresize}{!}{\includegraphics{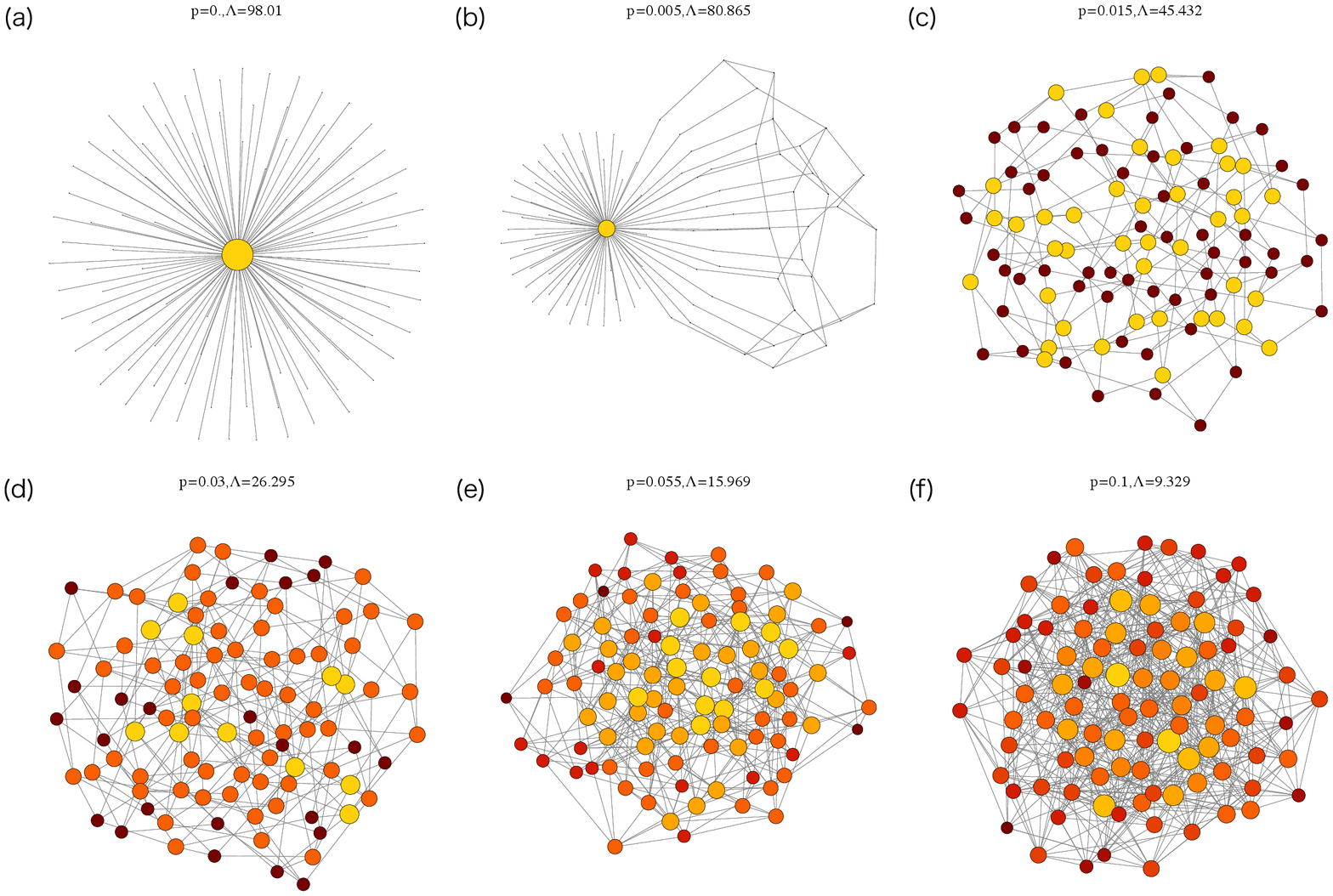}}
\end{center}
\vspace{-5mm}
\caption{(Color online). 
Synchronization-optimized networks with minimum $\Lambda$ values and $N=100$, for (a) $p=0.0$, (b) $p=0.005$, (c) $p=0.015$, (d) $p=0.03$, (e) $p=0.055$, and (f) $p=0.1$.
The nodes are colored according to degree, and their sizes are proportional to their degree.
The parameters are $M=8$ and $\kappa=1.0$.}
\label{fig:graph_large}
\end{figure*}

In Fig.~\ref{fig:graph_large}, the synchronization-optimized networks with the lowest $\Lambda$ values are shown for $N=100$.
Although the number of oscillator elements is larger,  we find the best synchronization-performance network is achieved for $p=0.0$, i.e., a star structure, using the REMC method (as shown in  Fig.~\ref{fig:graph_large}(a)).
When the connectivity, $p$, becomes slightly larger than zero, many loops appear, as shown in Fig.~\ref{fig:graph_large}(b).
At $p=0.015$, the synchronization-optimized network becomes homogeneous and all nodes are degree two or three [Fig.~\ref{fig:graph_large}(c)] and, as the connectivity increases,  ``small hubs'' start to appear and cores are formed, as shown in Fig.~\ref{fig:graph_large}(d)(e).
If $p$ converges to one, the network will be homogeneous again, because it approaches the complete graph. These results are consistent with the results 
obtained for a smaller system.

\subsection{Density of states and the number of connected labeled graphs}

\begin{table}[tdp]
\caption{The number of networks (states), $g(\Lambda)$, for a given synchronization performance, $\Lambda$, is estimated by WHAM. The results are in good agreement with the exact values, $g_{\mathrm{exact}}(\Lambda)$. The standard deviation errors estimated from 15 REMC trials are shown in parentheses and the parameters are $K=N-1, M=23$, and $\kappa=1.1$.}
\centering
\begin{tabular}{|c| | l | r | r |}\hline \hline
N   & $\Lambda$      & $g(\Lambda)$ & $g_{\mathrm{exact}}(\Lambda)$ \\  \hline \hline
     & $\Lambda_0\simeq 8.1$       & $9.96(0.06) \times 10^0$ & $10$ \\ 
     & $\Lambda_1\simeq 8.8$       & $7.18(0.02) \times 10^2$ & $7.20  \times 10^2$\\ 
10 & $\Lambda_2\simeq 9.3$       & $2.516(0.006) \times 10^3$ & $2.520 \times 10^3$ \\ 
     & $Z_{\star}$                            & $1.0001(0.0003)\times10^8$  & $1.000000  \times 10^8$\\ \hline
     & $\Lambda_0\simeq 13.067$  & $1.50(0.06)\times 10^1$ &  $15$ \\ 
     & $\Lambda_1\simeq 13.867$  & $2.730(0.011)\times 10^3$ & $2.730  \times 10^3$ \\ 
15 & $\Lambda_2\simeq 14.533$  & $1.638(0.007)\times 10^4$ & $1.6380  \times 10^4$ \\ 
     & $Z_{\star}$                             & $1.946(0.001) \times10^{15}$  & $1.9462\times10^{15}$ \\ \hline
     & $\Lambda_0\simeq 18.05$    & $2.00(0.02) \times 10^1$       & $20$ \\ 
     & $\Lambda_1\simeq 18.90$    & $6.82(0.07) \times 10^3$       & $6.840  \times 10^3$ \\ 
20 & $\Lambda_2\simeq 19.65$    & $5.80(0.05) \times 10^4$       & $5.8140  \times 10^4$\\ 
     & $Z_{\star}$                             & $2.622(0.007)\times 10^{23}$   & $2.62144\times10^{23}$ \\ \hline
     & $\Lambda_0\simeq 23.04$    & $2.51(0.07) \times 10^1$      & $25$ \\ 
     & $\Lambda_1\simeq 23.92$    & $1.39(0.03) \times 10^4$      & $1.3800  \times 10^4$ \\ 
25 & $\Lambda_2\simeq 24.72$    & $1.53(0.03) \times 10^5$      & $1.51800  \times 10^5$\\ 
     & $Z_{\star}$                             & $1.42(0.01) \times10^{32}$   & $1.42109\times10^{32}$ \\ \hline
     & $\Lambda_0\simeq 28.03$    & $2.98(0.11)  \times 10^1$      & $30$ \\ 
     & $\Lambda_1\simeq 28.93$    & $2.42(0.07) \times 10^4$       & $2.4360  \times 10^4$ \\ 
30 & $\Lambda_2\simeq 29.76$    & $3.29(0.01)  \times 10^5$      & $3.28860  \times 10^5$\\ 
     & $Z_{\star}$                             & $2.28(0.03)  \times10^{41}$   & $2.28768\times10^{41}$ \\ \hline \hline
\end{tabular}
\label{table}
\end{table}%

Before analyzing the statistical properties of the synchronization-optimized
 networks, we check whether REMC with WHAM can estimate the density of
 state
 accurately.
Although the number of combinations of labeled graphs with a given
$\Lambda$ is difficult to calculate in general, we can obtain 
the exact value for some $\Lambda$ when $K=N-1$. In this case, we can obtain
 $g_{\mathrm{exact}}(\Lambda_0)$, $g_{\mathrm{exact}}(\Lambda_1)$, and
 $g_{\mathrm{exact}}(\Lambda_2)$; the density
 of states for the smallest, second-smallest, and third-smallest $\Lambda$ values,
 exactly.
 We can also calculate $Z_{\star}$, the total number of connected
 networks, exactly. Using these values, we verify the performance of REMC
 with WHAM.

When the number of links is $K=N-1$, the graph exhibits a tree structure,
 the expected synchronization-optimized network is star shaped
 \cite{Deng09},
 and it has $\Lambda_0=\min_{g} \Lambda=(N-1)^2/N$. 
The number of possible combinations in the labeled star is $g_{\mathrm{exact}}(\Lambda_0)=N$, which corresponds to the number of ground states.
The graph with the second-smallest $\Lambda$ value, $\Lambda_1$, can be constructed by removing one of the periphery nodes in the star and attaching it to another periphery node.
Hence, the number of combinations of such labeled graphs is $g_{\mathrm{exact}}(\Lambda_1)=N (N-1)(N-2)$.
The graph having the third-smallest $\Lambda$ value, $\Lambda_2$, can be constructed by removing two periphery nodes from the star and attaching them to other periphery nodes.
Thus, the number of such combinations is $g_{\mathrm{exact}}(\Lambda_2)=n (n - 1) (n - 2) (n - 3)/2 $.
The exact value of $Z_{\star}$, the combinatorial of connected labeled graphs with a fixed
 number of links, is also known 
\footnote{The combinatorial of connected labeled graphs with a fixed number of links is shown in http://oeis.org/A123527}.
We can estimate the number of combinations for the labeled connected networks, $Z_{\star}$, with a fixed number of nodes and links through 
\begin{equation}
Z_{\star}=Z(\infty)-\sum_{\Lambda>\Lambda_{\max}} g(\Lambda),
\end{equation}
where 
\begin{equation}
  Z(\infty)=\frac{N!}{(N-K)!K!}\ ,
\end{equation}
is the total number of labeled graphs with $N$ nodes and $K$ links. We
check the accuracy of our method by comparing $g(\Lambda_0)$,
$g(\Lambda_1)$, $g(\Lambda_2)$, and $Z_{\star}$ as obtained by MCMC with the exact values. 
In Table~\ref{table}, we compare these exact values with the results estimated using REMC with WHAM. 
  These estimated values are in good agreement with the exact values,
  e.g., $Z_{\star}=2.28(0.03)  \times10^{41}$ for $N=30, K=29$, where the
  number in parentheses represents the error estimated by the standard
  deviation of 15 REMC trials.

\begin{figure}[tbp]
\begin{center} 
\resizebox{\singlefiguresize}{!}{\includegraphics{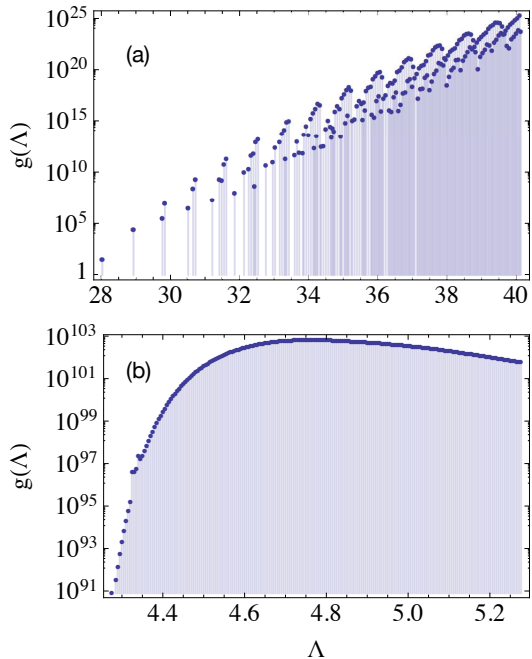}}
\end{center}
\vspace{-3mm}  
\caption{(Color online). Density of states estimated by WHAM for lower energy levels. (a) $K=29\; (p=0.0)$. The density of the lowest 200 energy levels is shown. (b) $K=110\; (p \sim 0.2)$. The density of the lowest 1,000 energy levels for every five levels is shown. The parameters are $N=30, M=23$, and $\kappa=1.2$.}
\label{fig:dos}
\end{figure}

In Fig. ~\ref{fig:dos}, we show the density (number) of states (graphs)
for $N=30$, $K=29$ and $N=30$, $K=110$.
Using the density of states, we calculate the statistical average of the synchronization-optimized networks, as will be discussed in later subsections.

\subsection{Statistical characteristics of the synchronization-optimized ensemble}

In this subsection, we analyze the statistical characteristics of
the synchronization-optimized ensemble, where the networks are sampled from
the canonical distribution defined in Eq.(\ref{eq:canonical}). In previous work, we
analyzed the degree distribution while, in this paper, we investigate
other properties of the synchronization-optimized ensemble. In the following, we present the results obtained from the ensemble at the highest inverse temperature, i. e.
 $\beta=\beta_M = \kappa^{M}-1$.

\begin{figure}[tbp]
\begin{center}
\resizebox{\singlefiguresize}{!}{\includegraphics{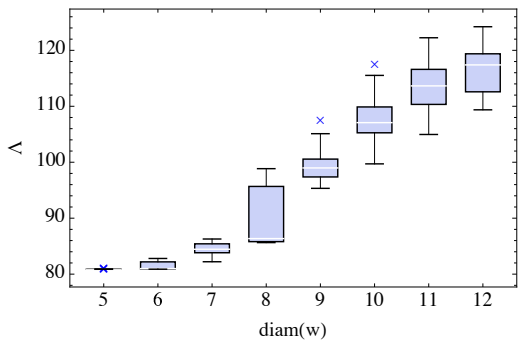}}
\end{center}
\vspace{-5mm}
\caption{(Color online). Distributions of $\Lambda$ for a fixed diameter, represented by a box-and-whisker chart. 
The crosses indicate outliers, and the 
parameters are $p=0.005, N=100, M=8$, and $\kappa=2.0$.}
\label{fig:char_dist_lambda}
\end{figure}

First, we investigate the relationship between the graph diameter and the
synchronization performance.
In Fig.~\ref{fig:char_dist_lambda},  box-and-whisker charts for the
distribution of  $\Lambda$ against a given diameter are shown.
 $\Lambda$ is determined by all the eigenvalues of the Laplacian matrix,
 and a small diameter does not always imply good synchronization performance.
This is easily seen from the fact that the $\Lambda$ values of a given network with
the same diameter are widely distributed, and there are overlaps in
the $\Lambda$ distributions for different diameters.
Nevertheless, Fig.~\ref{fig:char_dist_lambda} demonstrates that there is a
statistical tendency that the
synchronization performance increases as the diameter decreases in network ensembles that show good synchronization performance.

\begin{figure}[tbp]
\begin{center}
\resizebox{\singlefiguresize}{!}{\includegraphics{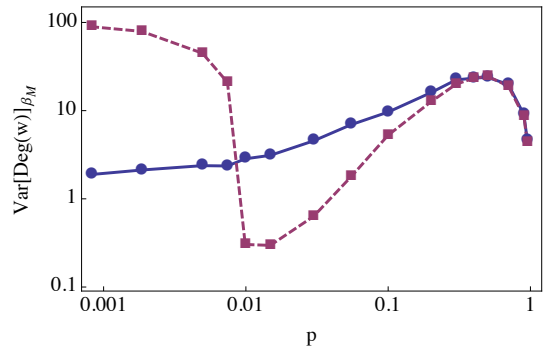}}
\end{center}
\vspace{-5mm}
\caption{(Color online). Degree distribution variance for synchronization-optimized and random networks as a function of connectivity, $p$. The blue solid and the red broken curves represent the random and the synchronization-optimized networks, respectively. The parameters are $N=100, M=23$, and $\kappa=1.2$.}
\label{fig:char_diameter_variance}
\end{figure}

Second, in Fig.~\ref{fig:char_diameter_variance}, we plot the
degree distribution variance for synchronization-optimized networks as a function of $p$.
In this figure, we also show the variance obtained for random networks in the interest of comparison. 
The dashed line shows the degree distribution variance for the
synchronization-optimized network ensemble, i.e., the average taken from
the many realizations, while the solid line represents the variance for random
networks, which are sampled by the replica with $\beta_0=0$.
We note that most of the networks created by the standard random network generator
 are disconnected for small $p$. Replica sampling is useful in avoiding this difficulty.
When $p=0$, the most synchronization-optimized network has a star structure, the
degree of center is $N$, and that of the other nodes is one.
Thus, the node degree variance of such star-like networks is approximately $N$.
Note that the synchronization-optimized ensemble is sampled at a finite
temperature (in the case of Fig.~\ref{fig:char_diameter_variance}, $\beta=255$), and thus the variance is the average over many optimized networks, almost all of which have star-like structures for $p=0.0$.
As the connectivity is slightly larger than zero, loops are formed (as shown
in Fig.~\ref{fig:graph_large}(b)), and the node degree variance gradually decreases.
The sudden decrease in the variance is observed in the vicinity of $p_c
\sim 0.02$, where the synchronization-optimized networks become the
most homogeneous [corresponding to the graph in
Fig.~\ref{fig:graph_large}(c), in which all nodes are degree two or three].
Above $p_c$, ``small hubs'' and cores start to appear, and thus the
variance for the synchronization-optimized network increases steadily, as
shown in Fig.~\ref{fig:char_diameter_variance}. 
As the connectivity increases further, the variance decreases, and the
difference between the synchronization-optimized and the random networks
becomes smaller, since both networks approach the complete graph.

\subsection{Anomaly in thermodynamic properties}

Thermodynamic properties are used to characterize
constraint-satisfaction problems such as the $N$-queens and Latin problems
\cite{Hukushima200277,PhysRevE.79.016703}. In these studies, the
 possibility of  phase-transition at system size $N\rightarrow \infty$ is investigated,
  however, no phase transition is found.
Hereafter, we consider the thermodynamic properties of the
synchronization-optimized networks, i.e., the statistical
characteristics of the ensemble of graphs defined in
Eq.~(\ref{eq:canonical}) depending on the inverse temperature, $\beta$,
 and investigate their dependence on $N$.

\begin{figure}[tbp]
\begin{center}
\resizebox{\singlefiguresize}{!}{\includegraphics{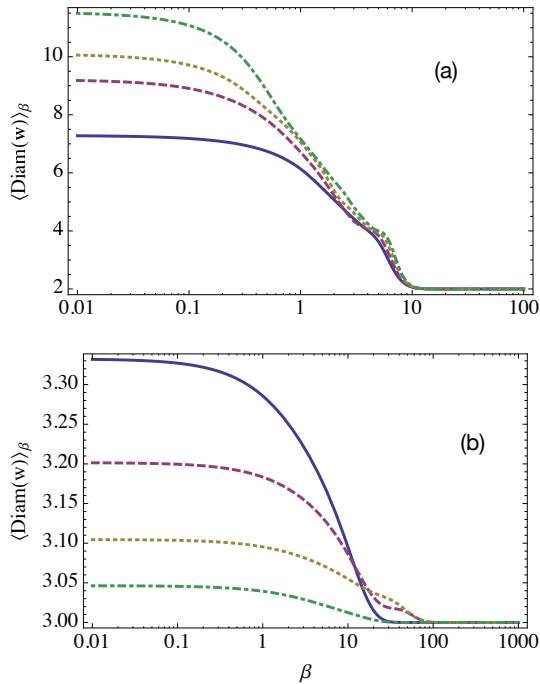}}
\end{center}
\vspace{-5mm}
\caption{(Color online).
Thermodynamic averages of diameters of synchronization-optimized networks are shown for (a) $p=0.0$ and (b) $p=0.05$.
The blue solid, red broken, yellow dotted, green dotted dash curves indicate the number of oscillators $N=15, 20, 25$, and $30$, respectively. The parameters are $M=16$ and $\kappa=1.3$.}
\label{fig:beta_diameter}
\end{figure}

In Fig.~\ref{fig:beta_diameter}, the thermal averages of diameters as a function of $\beta$ are shown for $p=0.0$ and $p=0.05$.
The thermal average of the diameter gradually decreases as $\beta$ increases, and converges to a
synchronization-optimized value.
For $p=0.0$, the value approaches $2.0$ as $\beta$ increases, and this is 
consistent with the fact that the most synchronization-optimized network
is star shaped.
Between the result for $p=0.0$ and $0.05$, a significant difference can
be seen in the system size dependence.
When $p=0.0$, the thermal average of the diameter for smaller $\beta$
increases linearly with the number of oscillators, $N$. 
On the other hand, when the number of links is slightly larger, i.e.,
$p=0.05$, the diameter of this random graph decreases as $N$ increases.
We also note that the slope of the curve in Fig.~\ref{fig:beta_diameter} becomes
smaller as $N$ increases.
Because we obtain a random network at $\beta=0$, this suggests that the difference in diameter between a random network and an optimized network decreases as $N\rightarrow \infty$, for $p=0.05$. 

\begin{figure}[tbp]
\begin{center}
\resizebox{\singlefiguresize}{!}{\includegraphics{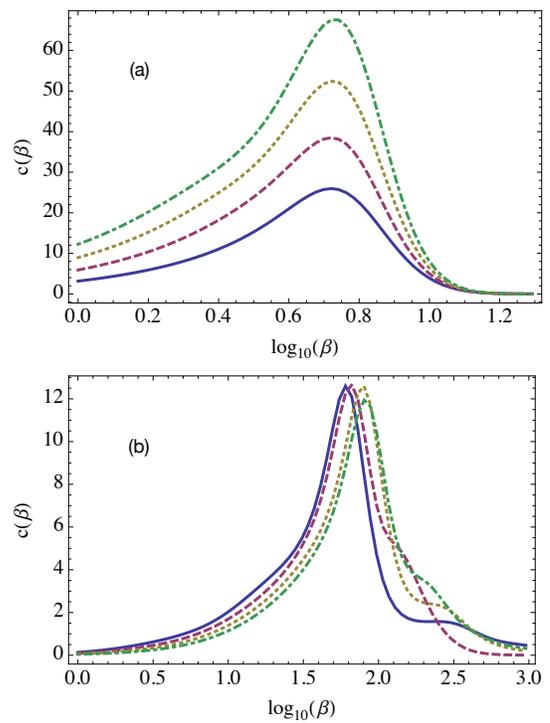}}
\end{center}
\vspace{-5mm}
\caption{(Color online). Heat capacity as a function of $\beta $ for (a) $p=0.0$ and (b) $p=0.20$.
The heat capacity diverges as the number of nodes increases, indicating that an anomaly exists in the vicinity of $\beta_c \sim 5.0$. The notations and parameters are the same as in  Fig.~\ref{fig:beta_diameter}.}
\label{fig:heat_capacity}
\end{figure}

Analogous to statistical mechanics, we show the heat capacity, $c(\beta)$, is
 defined by
 Eq.(\ref{eq:heat_capacity}), as shown in Fig.~\ref{fig:heat_capacity}.
We find an anomaly in the synchronization-optimized ensemble for smaller
 $p$, where the heat capacities (i.e., the synchronization performance fluctuation),
 show divergent behavior as the system size increases in the vicinity of
  $\beta_c \sim 5.0$. This is apparent in Fig.~\ref{fig:heat_capacity}(a).
By contrast, no divergent heat capacity behavior can be seen for the network with a larger number of links [Fig.~\ref{fig:heat_capacity}(b)].

An anomaly in heat capacity often implies a phase transition. However, the phenomenon in Fig.~\ref{fig:heat_capacity} can, in fact, be attributed to the divergence of the density of states and the Schottky anomaly \cite{Kubo86}.
As shown in Fig.~\ref{fig:dos}, there is a distinct gap in $\Lambda$ for
 $p=0.0$ and, in such systems, it is known that there exists a specific
heat phenomenon called the Schottky anomaly \cite{Kubo86}.
To demonstrate that the anomaly encountered in our result is indeed the Schottky anomaly, we
consider the two-level system in which $N$ atoms can have energy
$\epsilon_0$ or $\epsilon_0+\epsilon$.
Assuming that each energy level is $n_0$- and $n_1$- times degenerated, we obtain the heat capacity as
$$
c(T)=\frac{N \epsilon^2 \alpha  \; \exp(\epsilon/(k_B T))}{T^2
k_B( \exp(\epsilon/k_B T)+\alpha)^2},
$$ 
where $k_B$ is the Boltzmann constant and $\alpha=n_1/n_0$ is the degeneracy ratio
between the lower and higher energy levels.
This equation gives a peak in the specific heat, which is placed at
$\beta = \beta_s$, that satisfies
$$
(\beta_s \epsilon+2) \alpha=(\beta_s \epsilon-2)\exp(\beta_s \epsilon),
$$
and its height is given by
$$
\frac{1}{4}N k_B (\beta_s \epsilon - 2)(\beta_s \epsilon+2).
$$
From these equations, we find the following three important
observations. First, if $\alpha$ remains the same, $\beta_s$ and $\epsilon$ are
inversely proportional. Second, when $\alpha$ is
sufficiently large, $\beta_s$ increases logarithmically as $\alpha$
increases, because $(\beta_s-2)/(\beta_s+2)\sim 1$ in this limit.
Third, under the same assumption, the height of the heat capacity peak also
 increases as $ \beta_s^2 \epsilon^2 \sim \epsilon^2(\ln \alpha)^2$.

Our result is qualitatively consistent with that obtained by this simple
two-level model. When $p=0.0$, the energy gap between the ground state and
first excited state is $0.8$ for $N=15$, which increases slowly up to $1.0$ as
$N$ increases. On the other hand, the most optimized state and the second-most optimized state are degenerated $N$ and $N(N-1)(N-2)$ times,
respectively. Assuming that only these states contribute to the heat
capacity, the heat capacity peak is at $\beta\sim 7.39$ for
$N=15$, and 8.07 for $N=30$, while the observed peak exists at
$\beta\sim 5.0$. The height of the peak increases by more than three times
when $N=30$ compared with $N=15$, while the two-level approximation gives
the ratio 1.63. These values are slightly different from the result of
the numerical simulations, which is due to the fact that the 
less optimized states are neglected. Inclusion of these states will effectively
increase $\alpha$ and $\epsilon$, which results in a
decrease in $\beta_s$ and an increase in peak intensity. On the other hand, 
in the case of $p=0.20$, we have confirmed numerically that $\alpha$ is almost
 equal to 1 for any $N$. This is consistent with the independence of the peak
height from $N$ shown in  Fig.~\ref{fig:heat_capacity}(b). Therefore, our
analysis using this simple two-level model qualitatively explains the observed heat capacity anomaly
well.

\begin{figure}[tbp]
\begin{center}
\resizebox{\singlefiguresize}{!}{\includegraphics{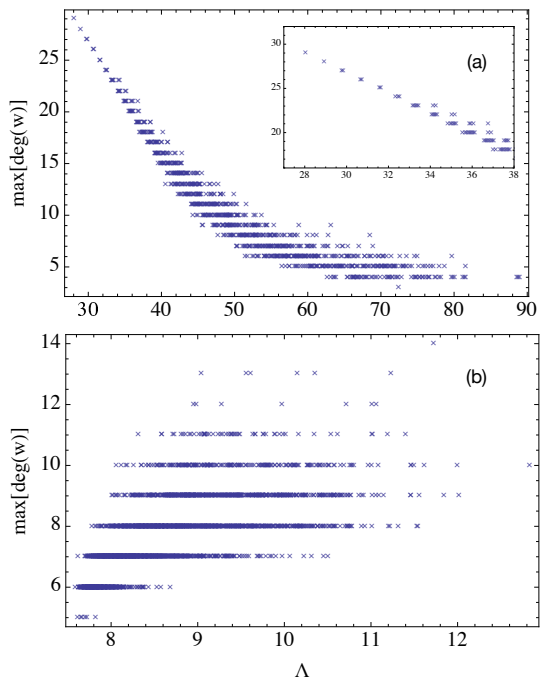}}
\end{center}
\vspace{-5mm}
\caption{(Color online). Scatter plot of maximum degrees of sampled networks against $\Lambda$.  
 (a) $p=0.0$ and (b) $p=0.10$, $N=30.$ 
The inset in (a) is a magnification and the other parameters are the same as for Fig.~\ref{fig:beta_diameter}.}
\label{fig:lambda_degree}
\end{figure}

In order to see the relationship between the network structure and the observed anomaly, a scatter plot of the maximum degree of sampled networks as a function of $\Lambda$ is given in  Fig.~\ref{fig:lambda_degree}.
This clearly shows that the synchronization performance decreases along with the degree of the center node for $p=0.0$, for which the network with the highest synchronization performance is a star-like network.
For greater connectivity, we see the synchronization performance decreases as the maximum degree increases, and networks with fixed $\Lambda$
 are widely distributed at the maximum degree. 
Furthermore, plateaus are apparent in Fig.~\ref{fig:lambda_degree}(a), and networks in the same plateau have star-like structures with the same degree as the center node, and have similar synchronization performance.
For cases of higher synchronization performance (smaller $\Lambda$ in the inset), there are gaps between the plateaus, and this corresponds to the energy gap in Fig.~\ref{fig:dos}.
Both the plateaus and the gaps simultaneously and gradually vanish with increasing connectivity. 

\section{Conclusions}
~\label{sec:summary}

We have considered an identical oscillator network under external
noise and designed a synchronization-optimized network using REMC.
To quantify the synchronization performance of the network, we have used
the Kirchhoff index, i.e., the sum of the inverse eigenvalues of the
Laplacian matrix for the network.
The density of states has been estimated using WHAM, i.e., the number of networks
for a given synchronization performance, and have obtained good
agreement with theoretical values.
Through the density of states, the statistical properties of the
synchronization-optimized networks have been analyzed.

The optimal network structure depends strongly on connectivity and, for smaller connectivity, the synchronization-optimized network has a star-like structure.
Such a network is inhomogeneous in a sense that the node
degree variance is larger, and it gradually changes to an almost homogeneous network
in the vicinity of $p = 0.015$. 
As the connectivity is further increased, hubs start to develop, and the
synchronization-optimized network becomes inhomogeneous.
As a result, the variance as a function of connectivity has two maxima
at $p=0.0$ (star) and $p \sim 0.2$ (core-periphery structure), and two
minima, $p \sim 0.015$ (homogeneous) and $p=1.0$ (complete graph). In general, there is a statistical tendency that the diameter of the network is related
to the synchronization performance. 

The change in structure in response to connectivity is not only apparent in the most
optimized network, but can also be seen in ensembles of well-optimized networks.
For example, the heat capacity of an ensemble of networks shows a divergent
peak as $N\rightarrow \infty$ when the connectivity is small, while it remains finite when the connectivity is large.
This anomaly is qualitatively explained by the increase in the degeneracy of the
excited state. Using a two-level approximation, we show that this
divergence can be accounted for by the Schottky anomaly with degeneracy.
In the case of small connectivity, the ratio of degeneracy between the first excited state
and the ground state increases rapidly, while it remains constant for large
connectivity.

These results provide us with useful information on network design for
synchronization.
For example, we have shown that, when the connectivity is small, all
well-optimized networks have star-like structures. However, a star-like
structure is vulnerable against attacks on the hub node. On the other
hand, when $p\sim 0.01$, the variance of the degree is at its smallest, and
the optimized network is homogeneous. We can expect that such a
homogeneous system will be capable of tolerating a targeted attack.
Therefore, a network with $p\sim 0.01$ will be preferable when a targeted attack
is suspected.
The statistical and thermodynamic properties determined in this study can therefore be applied to
the characterization of a wide range of problems, such as network optimization and
constrain-satisfaction problems.

\acknowledgments

This study has been partially supported by a Ministry of Education,
Science, Sports and Culture Grant-in-Aid for Scientific Research (Grant No. 24540417).


\bibliography{refkuramoto,refmcmc}
\end{document}